\documentclass{elsart}
\usepackage{natbib}
\usepackage{amsmath, amssymb, psfrag, graphicx, color, tabls}
\usepackage[latin1]{inputenc}

\newcommand{\vc}[1]{V\!(\text{C}|#1)} 
\newcommand{\vd}[1]{V\!(\text{D}|#1)} 
\newcommand{\id}[1]{\mathbf{1}_{(#1)}} 

\newtheorem{obs}[note]{Observation}

\journal{Journal of Theoretical Biology}

\begin{document}

\begin{frontmatter}

\title{Cooperation driven by mutations in multi-person Prisoner's Dilemma}

   \author{Anders Eriksson\corauthref{cor}}
   \ead{frtae@fy.chalmers.se}
   \author{Kristian Lindgren}
   \ead{frtkl@fy.chalmers.se}
   \corauth[cor]{Corresponding author. Phone: +46-31-772-3126. Fax: +46-31-772-3150}
   \address{Department of Physical Resource Theory, Chalmers University of Technology and Göteborg University, SE-41296 Göteborg, Sweden}

\begin{abstract}

\raggedright

The $n$-person Prisoner's Dilemma is a widely used model for
populations where individuals interact in groups. The evolutionary
stability of populations has been analysed in the literature for
the case where mutations in the population may be considered as
isolated events.
For this case, and assuming simple trigger strategies and many
iterations per game, we analyse the rate of convergence to the
evolutionarily stable populations. We find that for some values of
the payoff parameters of the Prisoner's Dilemma this rate is so
low that the assumption, that mutations in the population are
infrequent on that timescale, is unreasonable. Furthermore, the
problem is compounded as the group size is increased.
In order to address this issue, we derive a deterministic
approximation of the evolutionary dynamics with explicit,
stochastic mutation processes, valid when the population size is
large. We then analyse how the evolutionary dynamics depends on
the following factors: mutation rate, group size, the value of the
payoff parameters, and the structure of the initial population. In
order to carry out the simulations for groups of more than just a
few individuals, we derive an efficient way of calculating the
fitness values.
We find that when the mutation rate per individual and generation
is very low, the dynamics is characterised by populations which
are evolutionarily stable. As the mutation rate is increased,
other fixed points with a higher degree of cooperation become
stable. For some values of the payoff parameters, the system is
characterised by (apparently) stable limit cycles dominated by
cooperative behaviour. The parameter regions corresponding to high
degree of cooperation grow in size with the mutation rate, and in
number with the group size. For some parameter values, we find
more than one stable fixed point, corresponding to different
structures of the initial population.

\end{abstract}


\end{frontmatter}


\raggedright

\section{Introduction}
\label{sec:Introduction}

In social and natural systems, the action of an individual often
affects a number of other individuals, and there are numerous
examples of situations where so-called free riders or defectors
take advantage of others cooperating for a common good
\citep{hardin68,maynard_smith82,sugden86}.

A game-theoretic approach for the study of cooperation can be
based on the Prisoner's Dilemma game
\citep{flood58,rapoport_chammah65} -- a situation that captures
the temptation to act in a selfish way to gain a higher own reward
instead of sharing a reward by cooperating. In the game, the
players independently choose an action, either to defect or to
cooperate.

In the two-person game, the scores are $R$ (reward) for mutual
cooperation, $T$ (temptation score) for defection against a
cooperator, $S$ (sucker's payoff) for cooperation against a
defector, and $P$ (punishment) for mutual defection, with the
inequalities $S < P < R < T$ and (usually) $R > (T+S)/2$. We use
fixed values on R and S in this study, $R = 1$ and $S = 0$, while
$0 < P < 1 < T < 2$. (In the population dynamics we use there are
only three independent parameters, the third one being a growth
constant.)
From theoretic and simulation studies of two-person Prisoner's
Dilemma game, it is known under which circumstances repeated
interactions may allow for a cooperative population to be
established that can resist invasion by non-cooperative mutants
\citep[see, e.g.,][]{rapoport_chammah65, axelrod_hamilton81,
molander85, axelrod87, miller96, lindgren92, nowak_sigmund92}.

In the $n$-person Prisoner's Dilemma game, $n$ players
simultaneously choose an action, whether to cooperate or to
defect. In the literature, there are several evolutionary models
based on the $n$-person Prisoner's Dilemma using various strategy
sets and pairing mechanisms, e.g., where the players are
distributed in space. To our knowledge, the first evolutionary
models of this type are those by \citet{matsuo85} and
\citet{adachi_matsuo91, adachi_matsuo92}, where individuals
simultaneously play on a square lattice. A similar cellular
automaton was presented by \citet{albin92}.
In these models, a player performs a certain action every round
depending on previous actions of the nearest neighbours. Fitness
is given by joint actions of the player and its nearest
neighbours. This means that an action at a certain position may
influence the play at distant sites in the course of the iterated
game.
Models based on local groups on a lattice, where each player
participates in several overlapping but isolated games \citep[see,
e.g.,][]{matsushima_ikegami98, lindgren_johansson01}, can much
more easily exhibit a high degree of cooperation - the main
mechanism being a kind of kin or group selection allowing for the
formation of stable cooperative clusters, as has also been
observed for the 2-person game
\citep{nowak_may93,lindgren_nordahl94}.
\citet{hauert_schuster97} give a model involving probabilistic
strategies with finite memory. They find that an error-correcting
strategy type, previously studied in the context of the 2-person
Prisoner's Dilemma in \citep{lindgren92} and under the name
``Pavlov" in \citep{nowak_sigmund93}, may perform successfully
also in the multi-player case. Other extensions of the strategy
space introduce longer memory in terms of internal states
\citep[see, e.g.,][]{akimov_soutchanski97, lindgren_johansson01}.

In this paper, we revisit the classic $n$-person Prisoner's
Dilemma. Following \citet{boyd_richerson88} and
\citet{molander92}, the behaviours of the participants are
modelled by simple reactive strategies. These authors analyse the
stability of stationary populations in the limit where mutations
are infrequent: a mutation either is driven to extinction by the
selective pressure from the resident population, or leads to a new
resident population. The general conclusion from their studies is
that cooperation is difficult to obtain when extending the group
size beyond the two persons in the original Prisoner's Dilemma
game.
When mutations are infrequent, we find that for some values of the
payoff parameters, the rate of convergence to the evolutionarily
stable populations is so low that the assumption that mutations in
the population are infrequent on that timescale is unreasonable.
Furthermore, the problem is compounded as the group size is
increased. In order to address this issue, we derive a
deterministic approximation of the evolutionary dynamics with
explicit, stochastic mutation processes, valid when the population
size is large.
Among the questions we analyse are:
How does the evolutionary dynamics, and especially the long-term
average degree of cooperation, depend on how frequently mutations
occur in the population?
To what extent does the choice of parameter values in the
Prisoner's Dilemma payoff matrix affect the outcome?
How does the evolutionary dynamics depend on the group size?

The paper is organised as follows: In Section~\ref{sec:The
multi-person Prisoner's Dilemma game}, we define the multi-person
game and the strategy set used in our study. In
Section~\ref{sec:infinitesimal alpha}, the evolutionary dynamics
with an explicit flow of mutations is presented, and we review the
evolutionary stability analysis in the case of infrequent
mutations. In Section~\ref{sec:Evolutionary dynamics with
mutations}, we derive a deterministic approximation to the
evolutionary dynamics with mutations, and present the new method
for rapid calculation of fitness levels. In
Section~\ref{sec:Results}, the results from simulations and
numerical analysis of the evolutionary dynamics is presented and
discussed. In Section~\ref{sec:Summary and discussion} we conclude
with a summary and outlook.


\section{The $n$-person Prisoner's Dilemma game}
\label{sec:The multi-person Prisoner's Dilemma game}

In this section, we review the $n$-person Prisoner's Dilemma game
and define the strategy set used in our study. Each player
interacts with $n-1$ other players. Depending on the number $k$ of
others cooperating, a player receives the score $\vc{k}$ when
cooperating and the higher score $\vd{k}$ when defecting. The
scores increase with an increasing number of cooperators, and also
the total score given to all players increases if one player
switches from defection to cooperation, see, e.g.,
\citet{boyd_richerson88}. Thus
\begin{equation}\label{eq:payoff_func_conditions}
   \left\{
   \begin{array}{l}
      \vd{k} > \vd{k-1} \text{ and } \vc{k} > \vc{k-1} \\
      \vd{k} > \vc{k} \\
      (k + 1)\,\vc{k} + (n - k - 1)\,\vd{k + 1} >  \\\hspace{6cm} k\, \vc{k-1} + (n - k)\,\vd{k}.
   \end{array}
   \right.
\end{equation}
In this paper we shall assume that the scores $V$ can be
calculated as a linear combination of the scores against the other
players in $n-1$ ordinary two-player Prisoner's Dilemma games:
\begin{equation}
   \vc{k} = \frac{k}{n-1} \text{ and } \vd{k} = T\,\frac{k}{n-1} + P\,\frac{n - k - 1}{n-1} \, ,
\end{equation}
where we have divided by $n-1$ in order to make it easier to
compare results from different group sizes. The parameters $P$ and
$T$ obey $0 < P < 1 < T < 2$. Note that this is still an
$n$-person game since the same action is performed simultaneously
in all games.
This is a more general game than the one studied by
\citet{boyd_richerson88}, since they assume that $T - P = 1$
(corresponding to a line in our parameter region), but less
general than the payoff functions considered in
\citep{molander92}.
Other score functions that fulfil the conditions in
(\ref{eq:payoff_func_conditions}) are expected to yield similar
results \citep[c.f.][]{molander92}, and it is straightforward to
extend the the numerical methods in this paper to arbitrary score
functions $\vc{k}$ and $\vd{k}$, using the method for evaluating
the fitness values presented in Section~\ref{sec:Evolutionary
dynamics with mutations}.

It is assumed that the players have no way of coordinating their
choices, and that all player wish to maximise their score. Hence,
in the single round game, the optimal choice of action is to
defect, and this leads to all players in the group defecting and
scoring only $P < 1$ instead of 1, which they get if all
cooperate. If there is a high probability that the group will play
again, we have the iterated $n$-person Prisoner's Dilemma, and
then cooperation may develop under some circumstances.

We focus on the set of trigger strategies \citep{schelling78} as
the strategy space for the evolution, which was also considered
by, e.g., \citet{boyd_richerson88} and \citet{molander92}. A
trigger strategy $s_k$ is characterised by the degree of
cooperation that it requires in order to continue to cooperate: a
player with trigger strategy $s_k$ cooperates if at least $k$
other players cooperate. In a game with $n$ participants, $k$ is
in the range $0,\,\dots,\,n$. The strategy $s_0$ is an
unconditional cooperator and $s_n$ is an unconditional defector.
Each player decides whether to cooperate or to defect based on the
actions of the other players. In the first round after the
formation of a group, all players are assumed to cooperate, with
the exception of unconditional defectors. Then the players that
are unhappy with the number of cooperators switch to defection.
This may cause other players to change their action, and this is
iterated until a stable configuration has been reached. Note that
the number of cooperators may only decrease or be stable, and that
this procedure converges to the stable configuration with the
maximum number of cooperators. In a repeated game without noise,
this implies that a group of players with different trigger levels
reaches a certain degree of cooperation, some players may be
satisfied and cooperate while the others defect. Despite their
simplicity, trigger strategies capture many important aspects of
the many-person game, and allow for straight-forward evaluation of
the expected score for a player in a group randomly generated from
a given population.

It should be noted that some aspects of this game depend very
strongly on the expected number of iterations
\citep{boyd_richerson88}; most notably, the ability to defend
against pure defectors in large groups requires very long games,
to offset the cost of initial cooperative actions when defectors
are present in a group. In this article, however, we focus on the
role of mutations in the evolutionary dynamics. For simplicity,
and in order to treat the fitness values analytically, we limit
the study to the simpler situation when the games are infinitely
iterated, and we use the average score per player and round as a
fitness variable for the selection in the population dynamics.
This assumption can also be interpreted as a one-round game with
players making binding contracts in a negotiation before the game.


\section{Evolutionary dynamics}
\label{sec:infinitesimal alpha}

Consider a population of $N$ individuals. From one generation to
the next, a fraction $\delta$ of the population is replaced using
fitness proportional selection, where the fitness of an individual
is proportional to the number of offspring surviving to
reproductive age. Throughout this paper, $\delta = 0.1$. If small
enough, the value of $\delta$ does not influence the structure of
the evolving population, but determines the evolutionary time
scale. Assuming that the population size is large and constant,
the evolutionary dynamics takes the form of
\begin{equation}\label{eq:replicator_dynamics_simple}
   x'_l = x_l + \delta \left( \frac{f_l}{\bar{f}} - 1 \right) x_l \, ,
\end{equation}
where $x_l$ is the fraction of players in the population with
trigger level $l$, $x'_l$ is the value of $x_l$ in the next
generation, $f_l$ is the expected fitness for a player with
trigger level $l$, and $\bar{f} = \sum\limits_{l=0}^n x_l  f_l$ is
the average fitness in the population.

The expected fitness $f_l$ for a player with trigger level $l$ is
the expected score of the player in a randomly formed group:
\begin{equation}\label{eq:payoff}
    f_l = \sum\limits_{i_1,\, \dots,\, i_{n-1} =\, 0}^n
           x_{i_1} \cdots\ x_{i_{n-1}} \ S(l, i_1, \ldots ,i_{n-1})
\end{equation}
where $S(l, i_1, \ldots ,i_{n-1})$ is the score of a player with
strategy $s_l$ in a game with $n-1$ other players, using
strategies $s_{i_1},\, \ldots,\, s_{i_{n-1}}$ respectively.

\citet{molander92} gives an analysis of the model, with general
score functions $\vc{k}$ and $\vd{k}$, when it is assumed that a
mutation will either lead to a new resident population, or that
the evolutionary dynamics will bring the population back to the
situation before the mutation. Molander shows that in each
interval $P \in (\frac{k-1}{n-1}, \frac{k}{n-1})$, where $k \in
\{1,\dots,n-2\}$, there is either a mixture of strategies $s_k$
and $s_n$, which is evolutionarily stable, or there is a mixture
of strategies $s_0,\ldots,s_{n-1}$ (all cooperating), that resists
invasion by strategy $s_n$, but which is not evolutionarily
stable. Finally, there is no other asymptotically stable
population in that interval. In the interval $P \in
(\frac{n-2}{n-1},1)$, the purely cooperative equilibrium mixture
is the only possible asymptotically stable population.

\begin{figure}
\centerline{\includegraphics{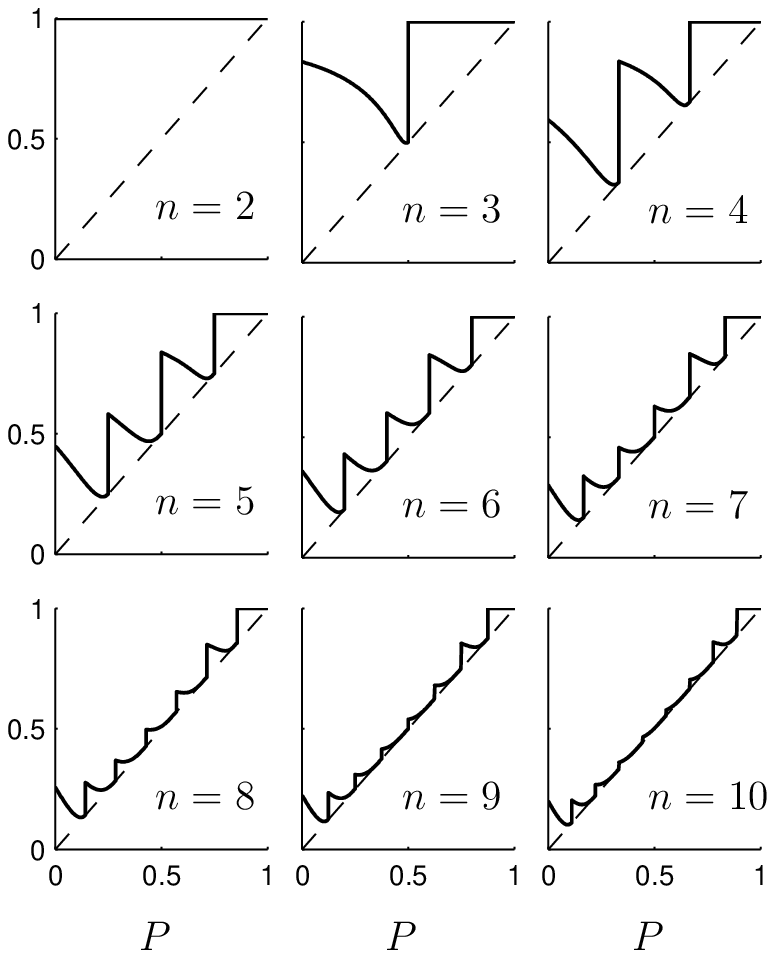}}

\caption{\label{fig:analytical_payoffs} The payoff at equilibrium
as a function of $P$ (thick line), for  $T = 1.2$ and group size
$n \in \{2,\dots,10\}$. Also shown is the payoff $P$ (dashed line)
for a population of pure all-defect.
}
\end{figure}

Suppose that a population with groups of size $n$ consists of a
mixture of strategies $s_k$ and $s_n$, in fractions $x$ and $1-x$,
respectively. Since, in this population, strategy $s_k$ cooperates
if and only if there are at least $k$ other players with the same
strategy in the group, and since strategy $s_n$ always defects,
direct evaluation of (\ref{eq:payoff}) gives
\begin{eqnarray}
    f_k(x) &=& \sum_{i=1}^k P {n-1 \choose i-1} x^{i-1} (1-x)^{n-i} \ +\nonumber\\
          && \hspace{3cm}+\ \sum_{i=k+1}^n \frac{i-1}{n-1} {n-1 \choose i-1} x^{i-1} (1-x)^{n-i}
\end{eqnarray}
for strategy $s_k$ and
\begin{eqnarray}
    f_n(x) &=& \sum_{i=0}^k P {n-1 \choose i} x^i (1-x)^{n-i-1} \ +\nonumber\\
          && \hspace{0cm}+\ \sum_{i=k+1}^{n-1}
          \left[P + (T - P) \frac{i}{n-1} \right] {n-1 \choose i} x^i (1-x)^{n-i-1},
\end{eqnarray}
for strategy $s_n$. We find the equilibrium by setting $f_k(x) =
f_n(x)$ and then solving for $x$, with the requirement $0 < x <
1$. Existence and uniqueness of this equilibrium is guaranteed by
the result of \citet{molander92}. In
Fig.~\ref{fig:analytical_payoffs} we show how the equilibrium
fitness depends on $P$ for $T = 1.2$, for group size $n \in
\{2,\dots,10\}$. Note that the fitness at the asymptotically
stable population approaches $f = P$ as $n$ increases, indicating
a decreasing degree of cooperation. From the existence and
uniqueness of the asymptotically stable populations of this form,
and from a Taylor expansion of $f_k$ and $f_n$ to the order
$x^{k+1}$, follows that $x \sim 1/n$ at the asymptotically stable
population, for $T > 1$.

\begin{figure}
   \centerline{\includegraphics{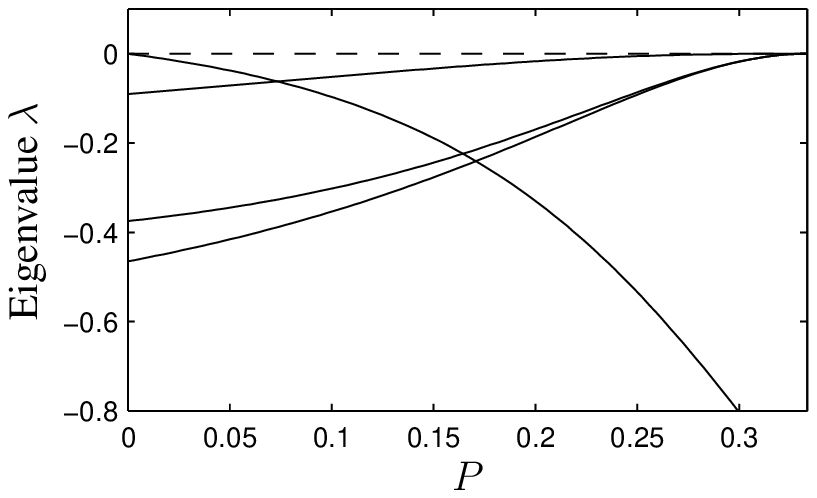}}
\caption{\label{fig:eig_ess_4}
Eigenvalues $\lambda$ of the Jacobian for the replicator dynamics
[c.f. (\ref{eq:replicator dynamics})], at the evolutionarily
stable population, in the limit of infrequent mutations. After a
small perturbation, the dynamics eventually returns to the
evolutionarily stable state, but the time it takes to do so
[characterised by $-(\ln 2)/\lambda$], diverges as $P$ approaches
$1/3$. The Jacobian, with elements $J_{ij} =
\frac{\partial}{\partial x_j}(f_i - \bar{f})$, is evaluated at the
stationary population, was obtained from explicit expressions of
$f_i$ and $\bar{f}$. These in turn were calculated using
(\ref{eq:payoff_l_final}) and Thm.~1.}
\end{figure}

Consider a small perturbation of one of the evolutionarily stable
populations constituted by the strategies $s_k$ and $s_n$. The
long-term response by the evolutionary dynamics to the
perturbation is given by the eigenvalues $\lambda$ of the Jacobian
of $x' - x$ at the evolutionarily stable population. We find that
the rate of convergence to the evolutionarily stable state
approaches zero, as $P$ approaches $k/(n-1)$, for $k \in
\{1,\dots,n-2\}$. This is illustrated in Fig.~\ref{fig:eig_ess_4}
for $n = 4$ and $P \in (0, 1/3)$. Thus, if $P$ is close enough to
$k/(n-1)$, the evolutionary dynamics does not relax to the
stationary population between mutations even for the case when
each perturbation is very small, provided that they occur
frequently enough. Since the selective forces are so weak, it
takes very long time to drive a mutation to extinction, and we
conclude that it is insufficient to analyse the evolutionary
properties in the limit of infrequent mutations. In the following,
we extend the evolutionary dynamics
(\ref{eq:replicator_dynamics_simple}) to explicitly include the
mutation process, and analyse how the evolutionary dynamics is
influenced by this process.

\section{Evolutionary dynamics with mutations}
\label{sec:Evolutionary dynamics with mutations}

We now introduce mutations as an explicit part of the evolutionary
dynamics. The population is subject to selection as in
(\ref{eq:replicator_dynamics_simple}). In addition, a number
$\mathcal{M}_{l \rightarrow j}$ of individuals per generation
switch from strategy $s_l$ to strategy $s_j$ due to mutations. The
mutations are assumed to be generated by stochastic events, e.g.
by a Poisson process, in the process of reproduction.
We assume that the probability of surviving to reproductive age
does not depend on the offspring's strategy, but only on that of
the parent. Hence, our measure of fitness comprises factors such
as parental care, as well as vying for resources.
It follows that the evolutionary dynamics takes the form
\begin{equation}\label{eq:replicator dynamics_stoch}
   x'_l = x_l + \delta \left( \frac{f_l}{\bar{f}} - 1 \right) x_l + \frac{1}{N} \sum_{j=0}^n ( \mathcal{M}_{j \rightarrow l} - \mathcal{M}_{l \rightarrow j} ).
\end{equation}
In a single generation, each player with strategy $s_l$ has an
expected number $\delta f_l x_l / \bar{f}$ of offspring surviving
to reproductive age. Mutations occur independently in the creation
of each offspring, with a probability of $\mu \ll 1$ per offspring
per generation, and the strategy of the mutated offspring is
chosen randomly among the $n + 1$ strategies (strategies $s_0,\,
\dots,\, s_n$), with equal probability. The expected number of
mutations per generation in the whole population is then $\mu\,N$.
Since the population size is assumed to be large, by the law of
large numbers \citep[see, e.g,][]{rice95} the number
$\mathcal{M}_{l \rightarrow j}$ of mutated offspring is
approximately equal to its expected value, given by:
\begin{equation}\label{eq:number_of_mutations}
   \mathcal{M}_{l \rightarrow j} \approx \frac{\mu\,\delta\,N\,f_l\,x_l}{(n+1)\,\bar{f}}.
\end{equation}
Note that this approximation is valid for all fitness values.
Inserting (\ref{eq:number_of_mutations}) into (\ref{eq:replicator
dynamics_stoch}), we obtain the following expression for the
evolutionary dynamics:
\begin{equation}\label{eq:replicator dynamics}
   x'_l = x_l + \delta \left( \frac{f_l}{\bar{f}} - 1 \right) x_l + \delta\,\mu \left( \frac{1}{n+1} - \frac{f_l}{\bar{f}}\,x_l \right).
\end{equation}
This equation is similar to the population dynamics introduced by
\citet{eigen71} in the study of quasi-species \citep[see
also,~e.g.,][]{eigen_schuster77,schuster_sigmund85}. Since $\mu
\ll 1$, the mutation process is only expected to have a
significant impact on the evolutionary dynamics when the fitness
values for all players are close to the average fitness in the
population, i.e. when the population is close to equilibrium.

We expect that the impact of a finite mutation rate on the
evolutionary dynamics is increasing with the population size.
Hence, the evolutionary dynamics of large groups are of special
interest. When the group size is small, or when only a couple of
strategies are present in the population (as in
Section~\ref{sec:infinitesimal alpha}), $f_l$ may be evaluated
from (\ref{eq:payoff}) by direct summation. For larger group
sizes, and when many strategies are present in the population,
this becomes impractical due to the rapid growth of the number of
terms in the sum, as shown in Table~\ref{tab:number of terms}.
Thus, in order to investigate the evolutionary dynamics of larger
groups, we need a way to evaluate the fitness values of players in
groups of more than just a few players.

\begin{table}
   \centerline{
   \begin{tabular}{rr}
   $n$ & Number of terms \\
   \hline
   2 & 9 \\
   3 & 64 \\
   4 & 625 \\
   5 & 7,776 \\
   6 & 117,649 \\
   7 & 2,097,152 \\
   8 & 43,046,721 \\
   9 & 1,000,000,000 \\
   10 & 25,937,424,601
   \end{tabular}}
\caption{\label{tab:number of terms} %
The number of terms in a direct evaluation of $f_l$ from
(\ref{eq:payoff}) for all $l$, for $n = 2\, \dots\, 10$. The
number grows very rapidly with $n$.
}
\end{table}

The population is characterised by the probability $P^l_i$ that
the number of cooperating players equals $i$, in a group with one
player using strategy $s_l$ and the $n-1$ other players chosen
randomly from the population. Since strategy $s_l$ cooperates if
$i > l$ we have
\begin{eqnarray}\label{eq:payoff_l_final}
    f_l = \sum_{i=0}^l P^l_i \, \vd{i} + \sum_{i=l+1}^n P^l_i \, \vc{i-1}.
\end{eqnarray}
Thus, efficient calculation of the $P^l_i$ allows for the study of
the evolutionary dynamics in larger groups. We derive a method for
calculating $P^l_i$, for arbitrary population compositions, that
avoids the rapid growth of the number of terms in
(\ref{eq:payoff}). Note that since $P^l_i$ only depends on the
distribution of trigger levels in the population, this method may
be applied for any payoff functions $\vc{i}$ and $\vd{i}$. Using a
simple observation on the number of cooperating players in a
group, we obtain:
\begin{thm}
In a population of trigger strategies, the probability $P^l_i$
that the number of cooperating players equals $i$, in a group with
one player using strategy $s_l$ and the $n-1$ other players chosen
randomly from the population, is
\begin{eqnarray}\label{eq:P forumla}
    P^l_i \ = \left\{
 \begin{array}{ll}
        0 & \text{ whenever } i = l, \\
        D^{l,0}_{n-1} & \text{ when } i = 0, \\
        \left( x_0 + \ldots + x_{n-1} \right)^{n-1} & \text{ when } i = n, \\
        \binom{n-1}{i - \id{l < i}}\
        \left( x_0 + \ldots + x_{i-1} \right)^{i - \id{l < i}}
        D^{l,i}_{n - 1 - i + \id{l \,<\, i}}
        & \text{ otherwise, }
    \end{array}
    \right.
\end{eqnarray}
where $\id{l \,<\, i}$ is one if $l < i$ and zero otherwise, and
$D^{l,i}_m$ is given by the recursive formula
\begin{eqnarray}\label{eq:D formula}
    D^{l,i}_m \ = \left\{
 \begin{array}{ll}
    x_n^m & \text{ when } i = n - 1 \\
    \sum\limits_{j = 0}^{M} \binom{m}{j}\, x_{i+1}^j\, D^{l,i+1}_{m - j} & \text{ otherwise}
    \end{array}
    \right.
\end{eqnarray}
where $M = m + i + 2 - \id{l \,<\, i} - n$.
\end{thm}
The proof is in the appendix. Note that whenever $l < i$ and $l' <
i$, $P^l_i = P^{l'}_i$, so if $l < i$ then $P^l_i  =  P^0_i$.
Using tables to store evaluated values of $D^{l,i}_m$, it is
possible to evaluate the values of $P^l_i$ for all $l$ and $i$ in
$\sim n^4$ operations. It may be possible to reduce this to $\sim
n^3$ operations through further exploitations of the recursive
structure of $D^{l,i}_m$.


\section{Results}
\label{sec:Results}

Armed with an efficient method for evaluating the fitness values,
we proceed to the results from our simulations. We have
investigated how the evolutionary dynamics [c.f.
(\ref{eq:replicator dynamics})] depends on different factors: the
payoff parameters $P$ and $T$, the initial population, group size
and mutation rate.

\subsection{Dependency on the payoff parameters}

We have measured the long-term degree of cooperation, over the
parameter region of $P$ and $T$, for various choices of group size
$n$. The initial state of the population is the always defecting
strategy $s_n$.

The long-term degree of cooperation is quantified by the
time-average and standard deviation of the average score in the
population, estimated from long simulations of the system, with
the initial transients removed. The standard deviation of
$\bar{f}(t)$ is defined as $\langle{\bar{f}(t)}^2\rangle -
\langle\bar{f}(t)\rangle^2$, where $\langle\cdot\rangle$ denotes
time average.

\begin{figure}
\centerline{
\begin{tabular}{c}
   \includegraphics{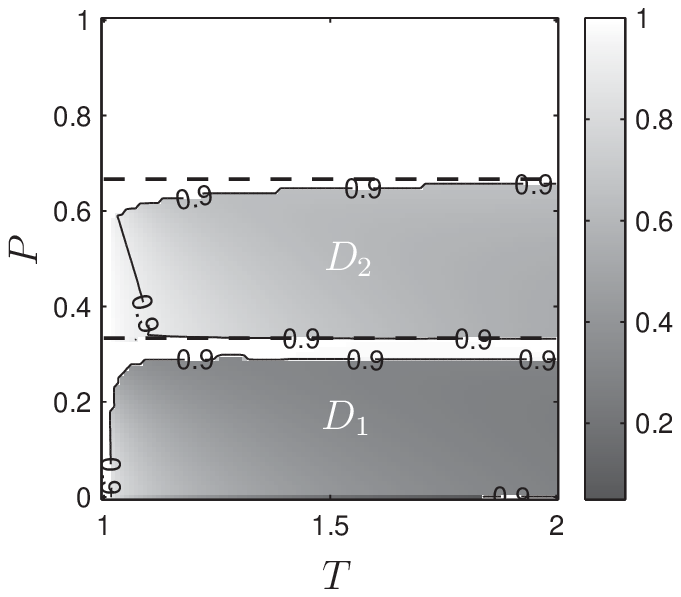} \\[5pt]
   \includegraphics{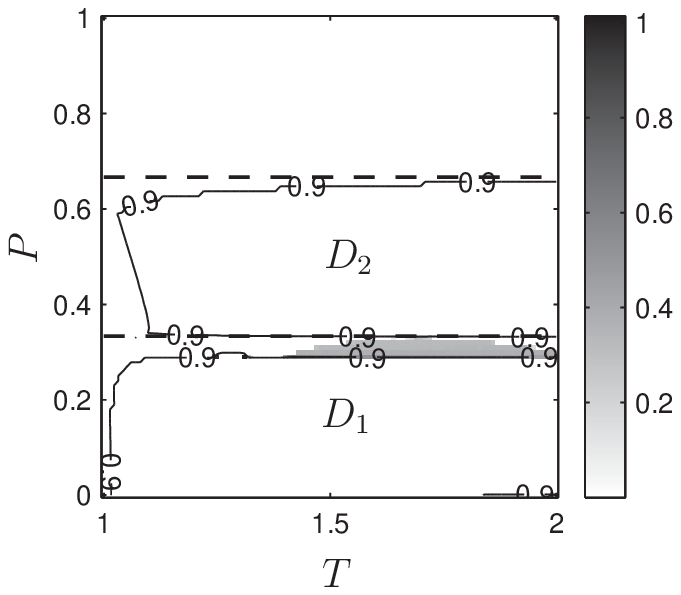}
\end{tabular}
}

\caption{\label{fig:favg n=4}
\textbf{Top} Asymptotic time-average of the average payoff in the
population (represented by the grey value, as indicated in the
scale to the right of the figure), as a function of the parameters
$T$ and $P$, for $n = 4$ and $\mu = 10^{-4}$. \textbf{Bottom}
Asymptotic standard deviation of the average payoff in the
population (represented by the grey value, as indicated in the
scale to the right of the figure), as a function of the parameters
$T$ and $P$. The averages and standard deviations were estimated
from long simulations of the system, with the initial transients
removed. The line where the average payoff equals $0.9$ is shown
in both panels. Also shown in both panels are the borders at $P =
1/3$ and $P = 2/3$ between the different asymptotically stable
populations, in the limit of infrequent mutations (dashed lines).
}
\end{figure}

In the top diagram of Fig.~\ref{fig:favg n=4}, the result for $n =
4$ is shown. For a large part of the domain, cooperation is high,
while there are two regions, $D_1$ and $D_2$, with significantly
lower payoffs. These two areas are characterised by fixed points
in the dynamics, dominated by strategy pairs ($s_1$, $s_4$) and
($s_2$, $s_4$), respectively.  These correspond to the fixed
points given by the evolutionary stability analysis, discussed in
Section~\ref{sec:infinitesimal alpha} and illustrated in
Fig.~\ref{fig:analytical_payoffs}. The other strategies are also
present, albeit at a much lower level (of the order of $\mu$): the
abundance of such a strategy is determined by a balance between
the net flow of mutations to the strategy, and the selective
pressure on the strategy from the population. In the limit $\mu
\rightarrow 0$, $D_1$ and $D_2$ corresponds to $0 < P < 1/3$ and
$1/3 < P < 2/3$, respectively, with $1<T<2$ in both cases.

In the lower part of Fig.~\ref{fig:favg n=4}, we give the standard
deviation of the average score (the transient excluded). A value
significantly above zero indicates a varying average fitness, such
as caused by the presence of oscillatory dynamics.
Note that most of the parameter region is dominated by stable
fixed points, corresponding to the fixed points of the dynamics in
the limit $\mu \rightarrow 0$. In the cooperative region, the
fixed points are given by a mixture of cooperating strategies
dominated by strategy $s_{n-1}$. An exception is the ridge between
region $D_1$ and $D_2$. In this region, stable fixed points are
characterised by a mixture of all strategies but the always
defecting strategy $s_n$, and thus a high level of cooperation,
contrary to what is expected from the stationary population in the
limit $\mu \rightarrow 0$. Furthermore, part of this area is
characterised by oscillating dynamics (as indicated by a non-zero
variance in the lower part of Fig.~\ref{fig:favg n=4}), between
cooperative and defecting strategies.

\subsection{Dependency on the initial population}

\begin{figure}
\centerline{
   \begin{tabular}{c}
      \includegraphics{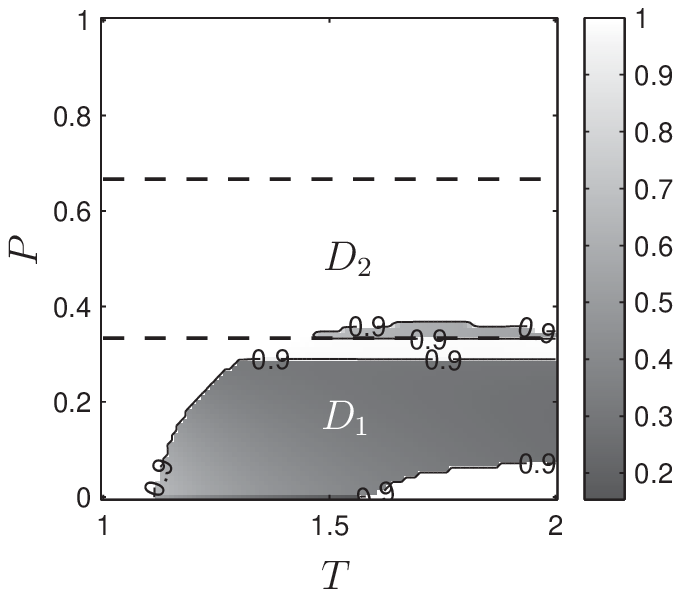} \\[5pt]
      \includegraphics{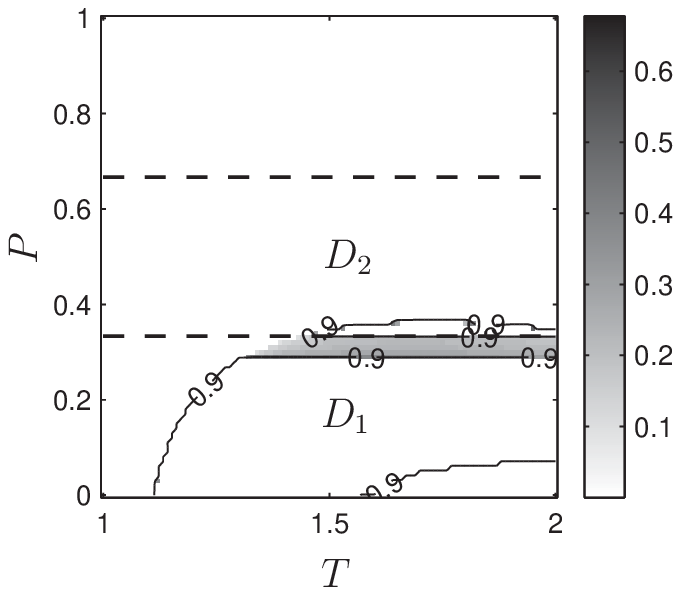}
   \end{tabular}
}

\caption{\label{fig:favg x3_n=4}
\textbf{Top} Asymptotic time-average of the average payoff in the
population (represented by the grey value, as indicated in the
scale to the right of the figure), as a function of the parameters
$T$ and $P$, when the population is initialised with a population
of $s_3$ strategies, in the case of group size $n = 4$ and $\mu =
10^{-4}$. \textbf{Bottom} Asymptotic standard deviation of the
average payoff in the population (represented by the grey value,
as indicated in the scale to the right of the figure), as a
function of the parameters $T$ and $P$. The line where the average
payoff equals $0.9$ is shown in both panels. Also shown in both
panels are the borders at $P = 1/3$ and $P = 2/3$ between the
different asymptotically stable populations, in the limit of
infrequent mutations (dashed lines).
}
\end{figure}

When the system is initialised with a population of $s_3$
strategies, in the case of group size $n = 4$, we find that in
some parts of the parameter region, there exist other stable fixed
points, see Fig.~\ref{fig:favg x3_n=4}. Most of region $D_2$ in
Fig.~\ref{fig:favg n=4} has been replaced by a cooperative fixed
point dominated by strategy $s_3$, and parts of region $D_1$ in
Fig.~\ref{fig:favg n=4} has been affected in a similar way. In
this case, cooperation is established and kept stable in the major
part of the parameter region.

In our model we have assumed that mutations are frequent in the
population, which alters the conclusions by Molander as discussed
in Section~\ref{sec:infinitesimal alpha} in two ways: First, the
cooperative region we observe for $P > 2/3$, is now a stable fixed
point, while there is no stable fixed point in that region in the
absence of mutations. Second, in the cooperative region between
regions $D_1$ and $D_2$, the flow of mutations destabilises the
fixed point given by the analysis of Molander discussed above,
resulting in either an oscillating distribution of strategies, or
a fixed point involving more than two strategies.

\subsection{Dependency on the mutation rate}

\begin{figure}
%
\centerline{
   \begin{tabular}{c}
      \includegraphics{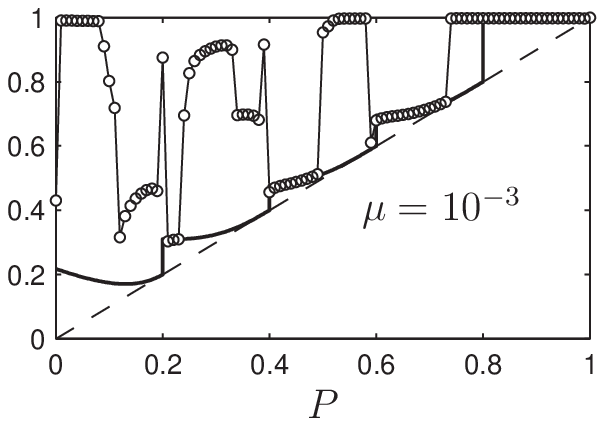}\\[5pt]
      \includegraphics{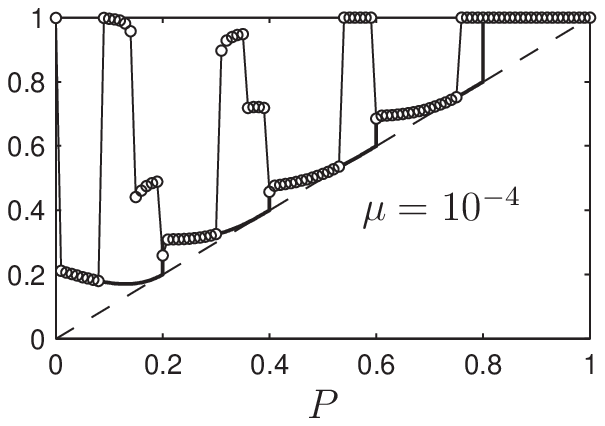}\\[5pt]
      \includegraphics{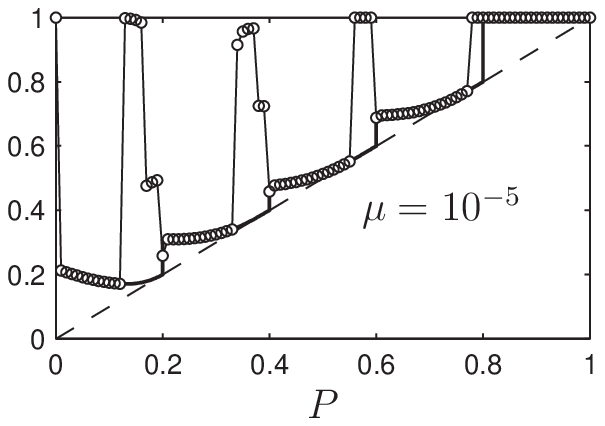}
   \end{tabular}
}

\caption{\label{fig:payoffs n = 6}
The time-averaged average payoff in the population as a function
of $P$, for $n = 6$ and $T = 1.6$, for $\mu = 10^{-3}$ (top), $\mu
= 10^{-4}$ (middle), and $\mu = 10^{-5}$ (bottom). Also shown is
the payoff in the limit of infinitesimal $\mu$ (thick line) and
the payoff $P$ for a population of pure all-defect (dashed line).
}
\end{figure}

\begin{figure}
\centerline{\includegraphics{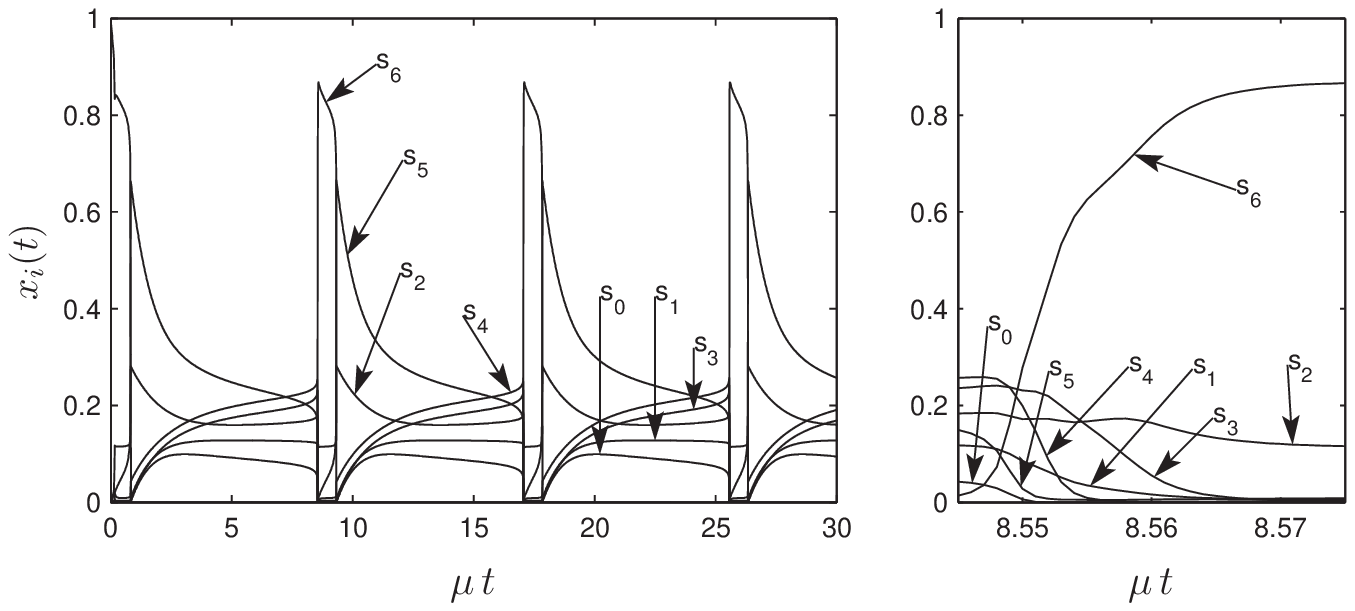}}

\caption{\label{fig:oscillations}
Time evolution of the population when $n = 6$, $\mu = 10^{-4}$, $T
= 1.6$ and $P = 0.33$. The left panel shows the initial transient,
and two cycles of the attractor. The right panel shows the
breakdown of cooperation, within a cycle, where strategy $s_6$
comes to dominate the population for a period of time. During the
part of the cycle where strategy $s_6$ is prominent, the system
approaches the population mixture which is evolutionarily stable
when mutations are infrequent. For the value of $P$ used here, it
corresponds to a mixture of strategies $s_2$ and $s_6$. For a
period of time, this mixture of strategies dominates the
population. After a while, however, $x_5$ starts to grow at the
expense of $x_6$, and after a while there is another sharp
transition where $x_5$ and $x_2$ grow and $x_6$ decay to very low
levels. The period of the oscillations is approximately
proportional to $(\delta\mu)^{-1}$.
}
\end{figure}

A more detailed numerical investigation for group size $n=6$ is
shown in Fig.~\ref{fig:payoffs n = 6}, where the average payoffs
are calculated as functions of $P$ for $T=1.6$. When increasing
the mutation rate (from $10^{-5}$ to $10^{-3}$), less of the fixed
points characterised by strategy pairs $(s_k, s_n)$ remains. Each
interval $(\frac{k-1}{n-1}, \frac{k}{n-1})$ of $P$ is
characterised by two regimes: At the lower end and in the middle,
the population is given by the $(s_k, s_n)$ equilibrium, indicated
by a thick line. At the higher end, the dynamics is either in a
stable limit cycle (see Fig.~\ref{fig:oscillations}) or in an
equilibrium involving more than two strategies, resulting in a
higher average score. In the higher end of these intervals, the
fraction of cooperating strategies vanish in the infinitesimal
mutation limit, so the effect of the finite mutation flow is that
cooperating strategies here gets an advantage. From the figure it
is clear that the regime for which the degree of cooperation is
high, grows with increased mutation rate.
The oscillatory dynamics (c.f. Fig.~\ref{fig:oscillations}) is
characterised by a period where the population is dominated by
cooperating strategies in the absence of the unconditional
defector, but a slow drift due to the mutation flow destabilises
that state. After a short interval in which the defector enters
but quickly disappears, the period starts again with the
cooperating strategies. The period of the oscillations is
approximately proportional to $(\delta\mu)^{-1}$.

\subsection{Dependency on the group size}

\begin{figure}
   \centerline{
   \begin{tabular}{c}
      \includegraphics{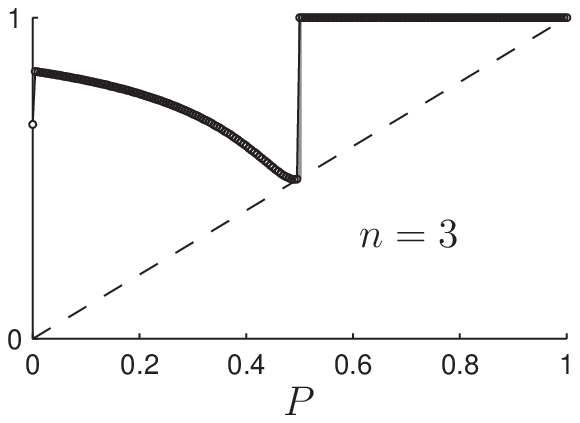}\\[5pt]
      \includegraphics{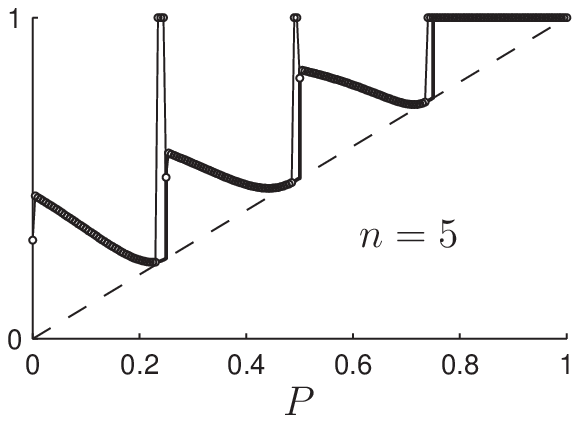}\\[5pt]
      \includegraphics{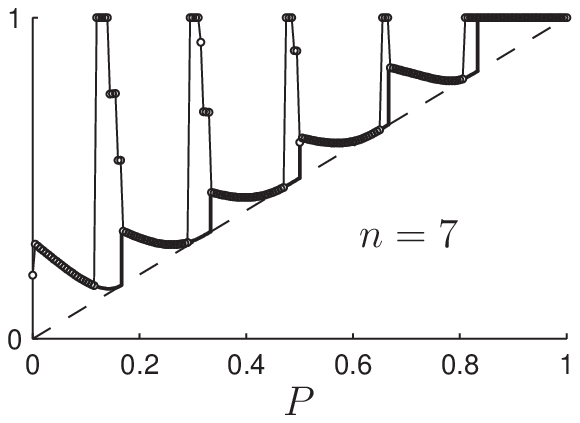}\\[5pt]
      \includegraphics{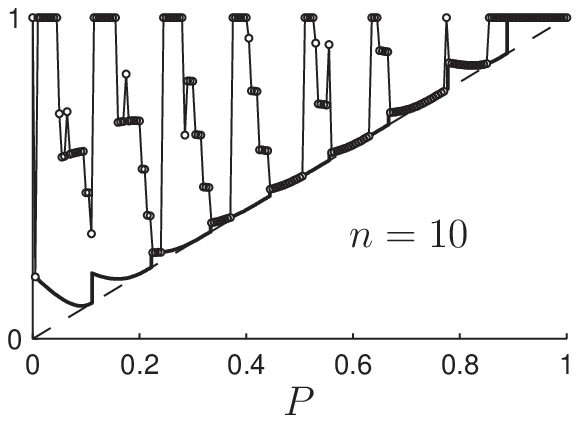}
   \end{tabular}
   }

\caption{\label{fig:payoffsdiff}
The time-averaged average payoff in the population for $n \in
\{3,5,7,10\}$ (top to bottom), for $T = 1.2$. Circles correspond
to simulations with $\mu = 10^{-7}$. Also shown is the payoff in
the limit of infinitesimal $\mu$ (thick line) and the payoff $P$
for a population of pure all-defect (dashed line).
}
\end{figure}

In Fig.~\ref{fig:payoffsdiff}, the average payoff as function of
$P$ is shown for four different group sizes, for $T = 1.2$. The
mutation rate is very low, $\mu = 10^{-7}$, and for group size $n
= 3$, the results are indistinguishable from the evolutionary
stability analysis based on fixed points involving only two
strategies. As the group size increases, we find that even for
this low mutation rate, there are significant differences compared
with the case of infinitesimal mutation flow. As the group size
gets larger, we get both an increasing number and size of
cooperative regions, with the exception of the original
cooperative region for $P > (n-2)/(n-1)$. This illustrates that we
can get a higher degree of cooperation when group size is
increased for some parameter values, while for other values
cooperation decreases.

As expected, the evolutionarily stable states are increasingly
sensitive to the rate of mutations, as the group size increases.
Also for very low rates of mutations, the average degree of
cooperation remains high as the group size increases, even though
the corresponding degree in the limit of infrequent mutations
converges to the payoff for a population of all-defect players.

\section{Summary and outlook}
\label{sec:Summary and discussion}

The $n$-person Prisoner's Dilemma game is generally assumed to be
characterised by a lower degree of cooperation than the two-person
game, when the players are confined to trigger strategies. We have
shown that this result is weakened when mutations occur frequently
in the population, even though the mutation rate per individual
per generation is low, and that it is due to the long relaxation
times of the evolutionary dynamics in certain areas of the
parameter region. This effect is increasing with group size, since
the number of different fixed point regions increases with number
of players. From this, we conclude that an analysis of the
evolutionary stability is not sufficient in order to achieve an
understanding of the long-term evolutionary dynamics.

In order to address this issue, we have derived a deterministic
approximation of the evolutionary dynamics with explicit,
stochastic mutation processes, valid when the population size is
large. We then analysed how the evolutionary dynamics depends on
the following factors: mutation rate, group size, the value of the
payoff parameters, and on the initial population. In order to
carry out the simulations for more than just a few individuals, we
have derived an efficient way of calculating the fitness values.

Our results may be summarised as follows: When the mutation rate
per individual and generation is very low, the dynamics is
characterised by populations that are evolutionarily stable. As
the mutation rate is increased, other fixed points with a higher
degree of cooperation become stable. For some values of the payoff
parameters, the system is characterised by (apparently) stable
limit cycles dominated by cooperative behaviour. The parameter
regions corresponding to high degree of cooperation grow in size
with the mutation rate and in number with the group size. For some
parameter values, there are more than one stable fixed point,
corresponding to different structures of the initial population.
When the group size is increased, the average degree of
cooperation in the population may increase or decrease depending
on exact payoff parameters and groups size.

The main limitation of this study is that the groups are assumed
to persist for a very long time. It is known that, e.g., the
ability to defend against pure defectors in large groups requires
very long games, to offset the cost of initial cooperative actions
when defectors are present in a group \citep{boyd_richerson88}. It
is thus of interest to extend the analysis of this paper to
investigate how the evolutionary dynamics is affected when the
expected number of iterations is varied.

In this article, we have assumed that all groups are of the same
size. In many situations, however, the group size can be expected
to vary from one group to another within the population. Our
results then indicate that a stable degree of cooperation would be
difficult to attain. However, if group size preferences would be
part of the players strategies, one could imagine that for certain
choices of parameters, the evolutionary dynamics may lead to the
formation of groups of certain sizes optimal for a high degree of
cooperation. In a forthcoming study, an analysis of dynamic group
formation in the $n$-person Prisoner's Dilemma will be presented.


\section*{Acknowledgements}

We thank the two anonymous referees for advice and comments.




{
\appendix
\section{Proof of Theorem 1}
\label{app:payoff derivation}

We begin with a simple observation on the number of players that
cooperate in the iterated game, when no player is unsatisfied with
the number of cooperators in the group. This is then exploited to
yield an efficient method of evaluating the probabilities $P^l_i$.

Let $k_j$ denote the number of players with trigger level $j$ in a
given group. Assume that strategies $s_0,\, \ldots,\, s_{i-1}$
cooperates, that strategies $s_i,\, \ldots,\, s_n$ defect, and
that no player will switch to defection in the next round. Since
strategy $s_{i-1}$ requires at least $i-1$ other players to
cooperate in the group, it cooperates in the next round if and
only if $\sum_{j=0}^{i-1} k_j \ge i$. Strategy $s_i$ defects in
the next round provided $\sum_{j=0}^{i-1} k_j \le i$. Thus, in a
stable configuration with $i$ cooperators, $\sum_{j=0}^{i-1} k_j =
i$ and $k_i = 0$. From this follows:
\begin{obs}\label{rem:stable configuration}
Assuming that all players cooperate in the first round, except
those with strategy $s_n$, the number $i$ of players that
cooperate in the stable configuration is determined by: (a) no
player has strategy $s_i$, (b) $i$ players have a trigger level
less than $i$, and (c) for the group there is no trigger level
larger than $i$ for which properties (a) and (b) hold.
\end{obs}

A randomly composed group with $n-1$ players contains $k_0 \ldots
k_n$ players, with trigger levels $0 \ldots n$ respectively, with
probability
\[
    (n - 1)!\ \frac{x_0^{k_0}}{k_0!} \cdots
     \frac{x_n^{k_n}}{k_n!}, \text{ where }
     \sum_{i=0}^n k_i = n - 1 .
\]
We now consider groups of size $n$ where one player has trigger
level $l$ and $n-1$ players are chosen randomly from the
population. Assume that strategies $s_0,\, \ldots,\, s_i$
cooperates, and that strategies $s_{i+1},\, \ldots,\, s_n$
defects. Then, in the next round
\[
   \left\{
   \begin{array}{rcl}
       \displaystyle
       0\ \ldots\ i \ \text{ cooperates }
       & \Leftrightarrow &
       \sum\limits_{j=0}^i k_j + \id{l \,\le\, i} \ge i+1 \\
       \displaystyle
       i+1\ \ldots\ n \ \text{ defects }
       & \Leftrightarrow &
        \sum\limits_{j=0}^{i+1} k_j + \id{l \,\le\, i+1} \le i+1
   \end{array}
   \right.
\]
where $\id{C}$ is one if the condition $C$ is true and zero
otherwise. We cannot have $l = i+1$ in a stable configuration,
since it would contradict Observation~\ref{rem:stable
configuration}, so $P^l_l = 0$ and we have
\begin{equation}\label{eq:fixpointcond_l}
    \text{$i$ is a stable configuration}\ \Rightarrow\
     \sum_{j=0}^{i-1} k_j  + \id{l < i} = i \text{ and } k_i = 0.
\end{equation}
This allows us to write $P^l_i$ as the product of two independent
sums:
\[
    P^l_i\ =\
     \sum\limits_{\substack{
        k_0,\, \dots,\, k_{i-1} \text{ s.t.}\\
        \text{$i$ is a stable conf.}
     }}
     \quad
     \sum\limits_{\substack{
        k_{i+1},\, \dots,\, k_n \text{ s.t.}\\
        k_0 +\, \dots\, + k_{n} = n \text{ and}\\
        \text{strategies $s_i,\ldots,s_n$ defects}
    }}
    \frac{n!}{k_0! \cdots k_n!}\ x_0^{k_0} \cdots\ x_n^{k_n} .
\]
The sum over $k_0\, \dots\, k_{i-1}$ evaluates to
\[
     \sum\limits_{\substack{
        k_0,\, \dots,\, k_{i-1} \text{ s.t.}\\
         k_0 + \dots + k_{i-1} = i - \id{l < i}
     }}\hspace{-5pt}
    \frac{x_0^{k_0}}{k_0!} \cdots \frac{x_{i-1}^{k_{i-1}}}{k_{i-1}!}
     = \
     \frac{1}{(i - \id{l < i})!}\ \left( x_0 + \ldots + x_{i-1} \right)^{i - \id{l < i}}.
\]
Unfortunately, the sum over $k_{i+1} \ldots k_n$ is not as simple
to evaluate. In order to evaluate it, we define $D^i_m$ to be the
sum over all groups where $m$ players with a trigger level greater
than $i$ is defecting (weighted by $m!$ to give a nicer result):
\[
    D^{l,i}_m = \sum\limits_{\substack{
        k_{i+1},\, \dots,\, k_n \text{ s.t.}\\
        k_{i+1} +\, \dots\, + k_n\ =\ m \text{, and} \\
        \text{strategies $s_i,\ldots,s_n$ defects given $l$}
    }}\hspace{-20pt}
    m!\
    \frac{x_{i+1}^{k_{i+1}}}{k_{i+1}!}
    \ \cdots\ \frac{x_n^{k_n}}{k_n!}
\]
By applying (\ref{eq:fixpointcond_l}) to $i \ldots n-1$ we see
that
\[
    D^{l,i}_m \ =
    \sum\limits_{\substack{
        k_{i+1},\, \dots,\, k_n \text{ s.t.}\\
        k_{i+1} +\, \dots\, + k_n\ =\ m \text{, and} \\
        k_{j+1} +\, \dots\, + k_n\ \ge\ n - j + \id{l \,<\, i} \text{ for} \\
        i\, \le\, j\, \le\, n-1
    }}\hspace{-20pt}
    m!\
    \frac{x_{i+1}^{k_{i+1}}}{k_{i+1}!}
    \cdots \frac{x_n^{k_n}}{k_n!} .
\]

The structure of the conditions on the sum allows us to express
$D^{l,i}_m$ recursively in terms of $D^{l,i+1}_m$, and we arrive
at (\ref{eq:P forumla}) and (\ref{eq:D formula}).

\hfill{} \qed

}

\end{document}